\documentclass[pre,floats,superscriptaddress]{revtex4}

\usepackage{graphicx}
\usepackage{epstopdf}
\usepackage{dcolumn}
\pdfoutput=1
\usepackage{amssymb}
\usepackage{amsmath,bm}
\usepackage{psfrag}
\usepackage{epsfig}
\usepackage{float}

\begin{document}
\title{Modeling solid-state dewetting of a single-crystal binary alloy thin films}

\author{Mikhail Khenner}
\affiliation{Department of Mathematics and Applied Physics Institute, Western Kentucky University, Bowling Green, KY 42101, USA}
\email{mikhail.khenner@wku.edu}

\begin{abstract}

\noindent
Dewetting of a binary alloy thin film is studied using a continuum many-parameter model that accounts for the surface and bulk diffusion, the bulk phase separation, the surface segregation and the particles formation.
Analytical solution is found for the quasistatic equilibrium concentration of a surface-segregated atomic species. This solution is factored into the nonlinear and coupled evolution
PDEs for the bulk composition and surface morphology. Stability of a planar film surface with respect to small perturbations of the shape and composition is analyzed, revealing the dependence of the 
particles size on major physical parameters. Computations show various scenarios of the particles formation and the redistribution of the alloy components inside the particles and on their surface. In most situations, for the alloy film composed initially of 50\% A and 50\% B atoms, a core-shell particles are formed, and
they are located atop a wetting layer that is modestly rich in the B phase. Then the particles shell is the nanometric segregated layer of the A phase, and the core is 
the alloy that is modestly rich in the A phase. 

\end{abstract}

\date{\today}
\maketitle


\section{Introduction}
\label{Intro}

Controlled solid-state dewetting, whereby a free-energy minimization drives the transformation of a continuous metal or semiconductor thin film into an ensemble of isolated particles, is emerging as a promising route to manufacture ordered micro- and nanoparticles arrays for new technologies based on nonlinear optics, plasmonics, photovoltaics, and photocatalysis \cite{Heger,Liao,Santbergen,Fafard}. Thus it is important to understand the physical mechanisms underlying such self-organization, and the past two decades have seen an acceleration
of the experimental and theoretical studies in this area.

Most experimental studies to-date of a solid-state dewetting of a \emph{binary alloy} thin films were conducted on polycrystalline films \cite{OKTR}-\cite{CZTB}. The primary pathway by which 
such film dewets is the nucleation and growth of holes at the grain boundary/substrate junctions \cite{SS1,SS2}. By contrast, a single-crystalline film may dewet either by a particle pinch-off from the retracting edge, or through a growth of a random surface defects \cite{JT,YT,YZSLLHL,SECS,BCLMOPL,AFRMB,DSMK,DOPL}. Either dewetting mode is sustained by a high-temperature surface diffusion of adatoms.  However, only a handful of studies exist that investigate the latter dewetting mode and the resultant formation of particles in the binary alloy films \cite{AKR}-\cite{AR}. These experiments were done on Fe-Au bilayer films heteroepitaxially grown on sapphire substrates
to the total thickness around 12nm, and then annealed at various temperatures in the interval 600$^\circ$C - 1100$^\circ$C (which are below the melting temperature of Fe and Au) for as long as 48 hrs to obtain the particles through dewetting.
Motivated by these studies, the goal of this paper is the development, analysis, and computation of a theoretical model that accounts for the major physical mechanisms that contribute to redistribution of the alloy components in the bulk of the film and on its surface, as the film undergoes dewetting and agglomeration into particles. No such model has been published, although there is an abundance of dewetting models for single-crystalline, single-component films (see, for instance, the review \cite{CLM} and the references therein).
In Ref. \cite{AKR1} the authors write: ``Very little is known about the kinetics of phase transformations in nano- and microparticles, both in terms of experimental observations and
at the level of available kinetic models". This paper partially fills the stated void in our understanding of how solid alloy particles are formed and what factors influence their compositions.

The proposed model is rooted in the sophisticated continuum (PDE-based) 
model by Zhang 
\textit{et al.} \cite{ZVD} of a core-shell nanowire growth; the prior versions of this model can be found in Refs. \cite{SVT,Tersoff,RS}. These earlier works consider a thick alloy film that is thermodynamically stable against phase separation. Also the thermodynamic surface segregation effects are not considered, and in Refs. \cite{SVT,Tersoff} the bulk diffusion is ignored. Another model worth mention is Ref. \cite{KB}, which adds the surface electromigration to the framework of Refs. \cite{SVT,Tersoff}. By contrast, the recent model \cite{ZVD} is thermodynamically consistent, thus it accounts for the surface and bulk diffusion, the surface and bulk phase separation (a thermodynamically unstable alloy is allowed), and for the thermodynamic and kinetic surface segregation. In the analysis of this model the authors focus on the effects of the surface diffusion and the kinetic surface segregation; the bulk diffusion and phase separation are given little attention.  In this paper the model of Ref. \cite{ZVD} is mapped onto a planar, substrate-supported film geometry, augmented by the physical boundary conditions, and a wetting/dewetting potentials (of van der Waals type) are introduced that sustain the film dewetting and its agglomeration into the particles. Despite the model complexity, we derive the closed system of two coupled nonlinear PDEs that is amenable for the analysis and computation. The outcome is the comprehensive and robust model that for the first time makes predictions on the coupled evolution of the solid film morphology and the film composition in the bulk and on the surface, as the film dewets and the particles are formed on the substrate. 

Our model assumes a continuous, large area solid alloy film of a nanoscale thickness. The initial condition of a film may be the surface corrugation (roughness) over the entire film area, or a localized surface defect (say, a pit), or a compositional non-uniformity/defect, or any combination of the above. We analyze the stability of a planar film surface with respect to
small perturbations of the shape and composition, and then 
compute the spatio-temporal evolution
of the initial condition. These numerical results are qualitatively matched to the experiments \cite{AKR}-\cite{AR}. Also, in order to elucidate the significant qualitative 
differences, where appropriate we compare the results to those from a simpler model for dewetting of a single-component film.

\section{Model formulation and the derivation of a coupled PDE system}
\label{Formulation}

\subsection{Model equations}

Due to high complexity of the model, we limit the consideration to 1D modeling. Thus all primary $(h,C_A,C_B,C_A^{(b)},C_B^{(b)})$ 
and secondary $(J_A,J_B,\mu_A,\mu_B,g^{(b)},\mu^{(b)},\gamma_A,\gamma_B,\gamma)$ variables are the functions of $x$ and $t$, where $x$ is the coordinate along the substrate,
and $t$ is the time. See the definitions of these variables below. We also employ the small-slope 
approximation (SSA). Thus $|h_x|\ll 1$ and the derivative with respect to arclength $S$ is approximated as $\partial/\partial S\approx \partial/\partial x$.
The $z$-coordinate is introduced in the direction perpendicular to the substrate and into the bulk of the film. $z=0$ corresponds to the substrate.

The model equations and boundary conditions are as follows.
\begin{itemize}

\item The conditions of the substitutional binary alloy for the surface and bulk concentrations of the atomic species A and B:
\begin{equation}
0< C_A(x,t),\ C_B(x,t),\ C_A^{(b)}(x,z,t), C_B^{(b)}(x,z,t) < 1,\quad C_A(x,t)+C_B(x,t)=1,\; C_A^{(b)}(x,z,t)+C_B^{(b)}(x,z,t)=1. \label{C_A_plus_C_B}
\end{equation}
These conditions will be used to eliminate $C_A$ and $C_A^{(b)}$.
The superscript $(b)$ here and elsewhere marks the
quantities in the bulk of a film.  The dimensionless concentrations are defined as follows. 
Let $\nu_0=\nu_A+\nu_B$ be the number density of the lattice sites occupied by A and B atoms on the film surface, where
$\nu_A$ and $\nu_B$ are the number of atoms of the components A and B per area on the surface.  
Then the surface concentrations are defined as $C_i=\nu_i/\nu_0,\; i=A,B$. Likewise, if $\nu_i^{(b)}$ are the number of atoms of the components A and B per volume in the bulk
and $\nu_0^{(b)}=\nu_A^{(b)}+\nu_B^{(b)}$ is the number density of the lattice sites in the bulk, then $C_i^{(b)}=\nu_i^{(b)}/\nu_0^{(b)}$.
\item The evolution PDE for the height $h(x,t)$ of the film surface:
\begin{equation}
V_n=h_t=-\Omega\left(\frac{\partial J_A}{\partial x}+\frac{\partial J_B}{\partial x}\right). \label{h_t_eq}
\end{equation}
According to Eq. (\ref{h_t_eq}), evolution of the surface is driven by the surface diffusion of different adatoms species A and B. $J_A$ and $J_B$ are the surface 
diffusion fluxes, $\Omega$ is the atomic volume, and $V_n$ is the normal velocity of the surface. 
\item The expressions for the surface diffusion fluxes:
\begin{equation}
J_i=-M_iC_i\frac{\partial \mu_i}{\partial x},\quad i=A\ ,B. \label{J_i}
\end{equation}
$M_A$ and $M_B$ are constant surface mobilities, and $\mu_A$, $\mu_B$ are the surface chemical potentials; they are shown 
below in Eqs. (\ref{mu_A}) and (\ref{mu_B}).
\item The evolution PDE for the surface concentration of B atoms:
\begin{equation}
\delta\frac{\partial C_B}{\partial t}+C_B^{(b)}V_n=-\Omega\frac{\partial J_B}{\partial x} +\Omega F_B. \label{C_B_eq}
\end{equation}
$\delta$ is the constant thickness of the surface layer and $F_B$ is the adsorption-desorption flux of B atoms on the bulk side of the surface. $F_B$ is shown below in Eq. (\ref{F_B}).
\item The evolution PDE for the bulk concentration of B atoms: 
\begin{equation}
\frac{\partial C_B^{(b)}}{\partial t}=-\Omega{\mathbf \nabla}\cdot {\mathbf G}_B^{(b)},\quad {\mathbf G}_B^{(b)}=-M_B^{(b)}{\mathbf \nabla}\mu^{(b)},
\quad {\mathbf \nabla}={\mathbf i}\frac{\partial}{\partial x}+{\mathbf k}\frac{\partial}{\partial z}. \label{C_B^b_eq}
\end{equation}
${\mathbf G}_B^{(b)}(x,z,t)$ and $M_B^{(b)}$ are the bulk flux and the constant atomic mobility in the bulk, respectively.
The expression for the bulk chemical potential $\mu^{(b)}$ is shown in Eq. (\ref{mu^b}) below.
\item The surface chemical potentials:
\begin{eqnarray}
\mu_A&=&g^{(b)}+\frac{C_B^{(b)}}{\delta}\left(\frac{\partial \gamma}{\partial C_A}-\frac{\partial \gamma}{\partial C_B}\right)-\gamma h_{xx} + 
\frac{\partial \gamma}{\partial h},\label{mu_A}\\
\mu_B&=&g^{(b)}-\frac{C_A^{(b)}}{\delta}\left(\frac{\partial \gamma}{\partial C_A}-\frac{\partial \gamma}{\partial C_B}\right)-\gamma h_{xx} + 
\frac{\partial \gamma}{\partial h}. \label{mu_B}
\end{eqnarray}
$g^{(b)}$ and $\gamma$ are the total bulk and surface energy, respectively, and $-h_{xx}$ is the surface curvature in the SSA.
The expressions for $g^{(b)}$ and $\gamma$ are shown below in Eqs. (\ref{gamma}) and (\ref{g^b}). Notice that the dependence of $\gamma$ on $h$ (the last term in Eqs. (\ref{mu_A}) and (\ref{mu_B})) is due to wetting/dewetting effect \cite{SZ,Chiu,Golubovic},
which is stated next in Eqs. (\ref{gamma}), (\ref{gamma_A}) and (\ref{gamma_B}).
\item The regular solution model for the total surface energy $\gamma$ in Eqs. (\ref{mu_A}) and (\ref{mu_B}):
\begin{equation}
\gamma = \gamma_A C_A+\gamma_B C_B+\alpha_{int}C_AC_B+kT\nu_0\left(C_A\ln C_A+C_B\ln C_B\right). \label{gamma}
\end{equation}
$\gamma_A$ and $\gamma_B$ are the surface energies of the alloy components, $\alpha_{int}$ is the interaction parameter on the surface 
(the surface enthalpy), and $kT\nu_0$ is the surface entropy (where $kT$ is Boltzmann's factor). 
\item The surface energies of the alloy components in Eq. (\ref{gamma}): 
\begin{eqnarray}
\gamma_A&=&\gamma_A^{(0)} + \frac{s^2p_2^{(A)}}{(h+s)^2}-\frac{sp_1^{(A)}}{h+s},\quad p_1^{(A)},\ p_2^{(A)} > 0, \label{gamma_A}\\
\gamma_B&=&\gamma_B^{(0)} + \frac{s^2p_2^{(B)}}{(h+s)^2}-\frac{sp_1^{(B)}}{h+s}, \quad p_1^{(B)},\ p_2^{(B)} > 0. \label{gamma_B}
\end{eqnarray}
The contributions $\gamma_A^{(0)}$ and $\gamma_B^{(0)}$ are the surface energies of a thick film $(h\rightarrow \infty)$ composed of either A or B atoms. The other two terms in either equation describe a wetting interaction of a film surface
with the substrate; they are non-vanishing when a film is thin. The terms $sp_1^{(i)}/(h+s)$ model the long-range van der Waals-type attraction, and the terms
$s^2p_2^{(i)}/(h+s)^2$ model the shorter-range repulsion, which results in a wetting layer of a thickness of the order of a wetting length $s$. 
Notice that $\gamma_i^{(0)}$ and the parameters $p_1^{(i)},\ p_2^{(i)}$ have the unit of energy per area.
This model and a similar models have been widely used, first for liquid films modeling and then for solid films modeling, 
see for instance Ref. \cite{Chiu} and Eq. (2.22) in Ref. \cite{Golubovic}. 
The parameters $p_1^{(i)}, p_2^{(i)}$ can be related to the familiar Hamaker constant $A_H$: $p\sim A_H/12\pi s^2$, where $A_H\sim 10^{-14} - 10^{-12}$ erg \cite{SZ},
and $s\sim 10^{-7}$ cm \cite{NMPBA,Chiu}. This gives $p_1^{(i)},\ p_2^{(i)} \sim 0.025 - 2.5$ erg/cm$^2$.  
\item The regular solution model for the total bulk energy in Eqs. (\ref{mu_A}) and (\ref{mu_B}):
\begin{equation}
g^{(b)}=g_AC_A^{(b)}+g_BC_B^{(b)}+\beta_{int}C_A^{(b)}C_B^{(b)}+kT\nu_0^{(b)}\left(C_A^{(b)}\ln C_A^{(b)}+C_B^{(b)}\ln C_B^{(b)}\right). \label{g^b}
\end{equation}
$g_A$ and $g_B$ are the bulk energies of the alloy components, $\beta_{int}$ is the interaction parameter in the bulk (the bulk enthalpy),
and $kT\nu_0^{(b)}$ is the bulk entropy.
\item The bulk chemical potential:
\begin{equation}
\mu^{(b)}=\frac{\partial g^{(b)}}{\partial C_A^{(b)}}-\frac{\partial g^{(b)}}{\partial C_B^{(b)}}. \label{mu^b}
\end{equation}
\item The adsorption-desorption flux on the bulk side of the film surface: 
\begin{equation}
F_B=-k_a\left[\frac{1}{\delta}\left(\frac{\partial \gamma}{\partial C_A}-\frac{\partial \gamma}{\partial C_B}\right)-\left(\frac{\partial g^{(b)}}{\partial C_A^{(b)}}-\frac{\partial g^{(b)}}{\partial C_B^{(b)}}\right)\right]. \label{F_B}
\end{equation}
$k_a$ is the adsorption-desorption coefficient.

\end{itemize}

Equations (\ref{C_A_plus_C_B})-(\ref{gamma}) and (\ref{g^b})-(\ref{F_B}) follow from Ref. \cite{ZVD} after the map onto a planar geometry and the application of SSA, and using a more convenient and standardized notations. Equations (\ref{gamma_A}) and (\ref{gamma_B}) read
$\gamma_A=\gamma_A^{(0)}(\theta)$, $\gamma_B=\gamma_B^{(0)}(\theta)$ in Ref. \cite{ZVD}, where the right-hand sides
are certain functions of the surface orientation angle $\theta$; this, and the inclusion of the anisotropy terms in the chemical potentials $\mu_A$ and $\mu_B$ introduce the anisotropy into the model of Ref. \cite{ZVD}. Due to abundance of other important physical effects and the associated parameters whose role is seldom brought to light and thus remains unclear, in this paper we choose not to study the anisotropy effects, thus $\gamma_A^{(0)}$ and $\gamma_B^{(0)}$ are constants. Also, the model of Ref. \cite{ZVD} includes the deposition of atoms on the film surface from a beam or from a vapor phase. In this model these contributions are omitted, since the post-deposition film evolution under annealing is of interest.

The boundary conditions at the surface and at the substrate follow from mass conservation:
\begin{eqnarray}
z=h:\quad {\mathbf n}\cdot{\mathbf G}_B^{(b)}&=&F_B, \label{bc_surface}\\
z=0: \quad \frac{\partial \mu^{(b)}}{\partial z}&=&0. \label{bc_substrate}
\end{eqnarray}
In Eq. (\ref{bc_surface}) ${\mathbf n}$ is the unit normal to the surface, pointing into the vapor phase.  Substituting ${\mathbf G}_B^{(b)}$ from Eq. (\ref{C_B^b_eq}) into the boundary condition (\ref{bc_surface}) and applying SSA gives the final form of the boundary conditions:
\begin{eqnarray}
z=h:\quad M_B^{(b)}\frac{\partial \mu^{(b)}}{\partial z}&=&M_B^{(b)}h_x \frac{\partial \mu^{(b)}}{\partial x}-F_B, \label{bc_surface1}\\
z=0: \quad \frac{\partial \mu^{(b)}}{\partial z}&=&0. \label{bc_substrate1}
\end{eqnarray}

At first glance, after $C_A$ and $C_A^{(b)}$ are eliminated using Eqs. (\ref{C_A_plus_C_B}), the system (\ref{C_A_plus_C_B})-(\ref{F_B}), (\ref{bc_surface1}), (\ref{bc_substrate1})
should reduce to three coupled, very nonlinear evolution PDEs for $h$, $C_B$, and $C_B^{(b)}$, plus the boundary conditions.  
Computing the numerical solution of that system would be difficult since each of the three variables must be discretized separately, resulting in tripling of the number of the grid unknowns.
This problem is compounded by the nature of the dewetting model, which typically requires a fine grid to resolve the particles.  
Thus our approach is through the problem reduction. As is totally expected and natural, the reduction and simplification come at a cost of neglecting certain 
secondary physical effects. Precisely, we show below that assuming $\alpha_{int}=0$ in Eq. (\ref{gamma}) allows to eliminate $C_B$; alternatively, assuming $\beta_{int}=0$ in Eq. (\ref{g^b}) allows to eliminate $C_B^{(b)}$. Thus a much simpler problem for only two coupled
variables ($h, C_B$, or $h, C_B^{(b)}$) can be obtained. We will choose to eliminate $C_B$ by assuming $\alpha_{int}=0$. This assumption is less restrictive than the assumption
$\beta_{int}=0$. It means, in particular, the impossibility of an independent from the bulk phase separation at the surface. Notice that
the possibility of the bulk phase separation is retained through $\beta_{int}\neq 0$ and, if there is such phase separation, then it  
drives the one at the surface.

\subsection{Derivation of the PDEs for $h$ and $C_B^{(b)}$}

We start the solution of the system (\ref{C_A_plus_C_B})-(\ref{F_B}), (\ref{bc_surface1}), (\ref{bc_substrate1})  from Eq. (\ref{C_B^b_eq}). Substituting ${\mathbf G}_B$ and integrating from $0$ to $h$ gives
\begin{equation}
\int_0^h \left(\frac{\partial C_B^{(b)}}{\partial t}-\Omega M_B^{(b)}\frac{\partial^2 \mu^{(b)}}{\partial x^2}\right)dz = \Omega M_B^{(b)} \frac{\partial \mu^{(b)}}{\partial z}|_{z=h} - \Omega M_B^{(b)} \frac{\partial \mu^{(b)}}{\partial z}|_{z=0}.
\end{equation}
Applying the boundary conditions (\ref{bc_surface1}) and (\ref{bc_substrate1}) gives
\begin{equation}
h \frac{\partial C_B^{(b)}}{\partial t} = \Omega M_B^{(b)} \frac{\partial}{\partial x}\left(h \frac{\partial \mu^{(b)}}{\partial x}\right) -\Omega F_B.
\label{e1}
\end{equation}
Next, we substitute $V_n$ from Eq. (\ref{h_t_eq}) into Eq. (\ref{C_B_eq}), find $\Omega F_B$ from the latter equation and substitute that expression in Eq. (\ref{e1}):
\begin{equation}
h \frac{\partial C_B^{(b)}}{\partial t} = \Omega M_B^{(b)} \frac{\partial}{\partial x}\left(h \frac{\partial \mu^{(b)}}{\partial x}\right) -
\delta\frac{\partial C_B}{\partial t}+\Omega C_B^{(b)}\left(\frac{\partial J_A}{\partial x}+\frac{\partial J_B}{\partial x}\right)-\Omega\frac{\partial J_B}{\partial x}.
\label{C_B^b_t_eq1}
\end{equation}

Notice that (i) Eq. (\ref{C_B^b_t_eq1}) contains the time derivatives of $C_B^{(b)}$ and $C_B$, and (ii) $F_B$ (see Eq. (\ref{F_B})) is not known, and it has been eliminated from the equations and boundary conditions of the system. Also, Eq. (\ref{h_t_eq}) is the evolution PDE for $h$. With the help of Eq. (\ref{F_B}) we will express $C_B$ through $C_B^{(b)}$ and thus Eq. (\ref{C_B^b_t_eq1}) will turn into an evolution PDE for $C_B^{(b)}$. To this end we introduce the constitutive relation that states the weights of two contributions to $F_B$ on the right-hand side of Eq. (\ref{F_B}):
\begin{equation}
\frac{1}{\delta}\left(\frac{\partial \gamma}{\partial C_A}-\frac{\partial \gamma}{\partial C_B}\right)=\xi \left(\frac{\partial g^{(b)}}{\partial C_A^{(b)}}-\frac{\partial g^{(b)}}{\partial C_B^{(b)}}\right).
\label{from_F_B}
\end{equation} 
Here $\xi$ is the positive and dimensionless proportionality parameter. Taking $\xi=1$ corresponds to fast adsorption-desorption kinetics, such that $F_B/k_a\approx 0$. This is
the McLean's condition \cite{McLean}. In this paper we use $0<\xi\le 10$, with most computations done at the McLean's model value $\xi=1$. 
Substituting Eqs. (\ref{gamma}) and (\ref{g^b}) in Eq. (\ref{from_F_B}), we easily find:
\begin{equation}
\frac{kT\nu_0}{\delta}\ln{\frac{1-C_B}{C_B}}+\frac{2\alpha_{int}}{\delta}C_B=\xi\left[k T\nu_0^{(b)}\ln{\frac{1-C_B^{(b)}}{C_B^{(b)}}}+2\beta_{int}C_B^{(b)}+g_A-g_B-\beta_{int}\right]-
\frac{1}{\delta}\left(\gamma_A-\gamma_B\right).
\label{from_F_B_1}
\end{equation}
As discussed above, we set $\alpha_{int}=0$ in Eq. (\ref{from_F_B_1}), which yields the analytical solution of that equation:
\begin{equation}
C_B\left(C_B^{(b)}\right)=\left[\left(\frac{1}{C_B^{(b)}}-1\right)^{\frac{\xi\delta\nu_0^{(b)}}{\nu_0}}+\exp{\left(\frac{2\xi\delta\beta_{int}}{kT\nu_0}C_B^{(b)}\right)}+1+\Psi\right]^{-1},
\label{C_B_through_C_B^b}
\end{equation}
where
\begin{equation}
\Psi= \exp{\left(\frac{\delta}{kT\nu_0}\left[\xi\left(g_A-g_B-\beta_{int}\right)-\frac{1}{\delta}\left(\gamma_A^{(0)}-\gamma_B^{(0)}\right)\right]\right)}
\label{Psi}
\end{equation}
does not depend on $C_B^{(b)}$.
According to Eq. (\ref{C_B_through_C_B^b}), the surface concentration of B atoms is determined by the bulk concentration of B atoms, the bulk and surface energies of A and B atoms, the bulk enthalpy, the temperature, and the phenomenological parameter $\xi$.
From the form of this equation it can be noticed that $0<C_B<1$, as expected. This form 
again reminds of the
McLean model \cite{McLean,WK,S}, in which the surface concentration $X_s$ of the component B and the concentration $X_b$ of this component in the bulk region adjacent to the surface are related as follows: 
$X_s=\left[\left(\frac{1}{X_b}-1\right)\exp{\left(\Delta H/RT\right)}+1\right]^{-1}$, where $\Delta H$ is the enthalpy of segregation of B at the surface.
In fact, Eq. (\ref{C_B_through_C_B^b}) can be seen as the generalization of the McLean model to the bulk phase-separating alloy film with the thermodynamically stable surface.
To simplify equations writing in the remainder of this Section, in Eq. (\ref{Psi}) we replaced $\gamma_i$ by $\gamma_i^{(0)}$, since at our choice
of the same value for $p_1^{(i)}$ and $p_2^{(i)}$ (see Table \ref{table1}) the fractions containing $h$ cancel out when $\gamma_B$ is subtracted 
from $\gamma_A$ (see Eqs. (\ref{gamma_A}) and (\ref{gamma_B})).

To derive the final form of the evolution PDE for $C_B^{(b)}$, we substitute Eqs. (\ref{J_i}) and (\ref{C_B_through_C_B^b}) in Eq. (\ref{C_B^b_t_eq1}), perform differentiation
of $C_B$ using the Chain Rule, use conditions (\ref{C_A_plus_C_B}) and obtain the following equation:
\begin{equation}
\left[h+\delta \Phi\right]\frac{\partial C_B^{(b)}}{\partial t}=\Omega M_B^{(b)} \frac{\partial}{\partial x}\left(h \frac{\partial \mu^{(b)}}{\partial x}\right)-
\Omega M_A C_B^{(b)} \frac{\partial}{\partial x}\left[\left(1-C_B\right) \frac{\partial \mu_A}{\partial x}\right]+
\Omega M_B \left(1-C_B^{(b)}\right) \frac{\partial}{\partial x}\left[C_B \frac{\partial \mu_B}{\partial x}\right].
\label{C_B^b_t_eq_final}
\end{equation}
Here $\mu_A,\ \mu_B,\ \mu^{(b)}$ are given by Eqs. (\ref{mu_A}), (\ref{mu_B}), (\ref{mu^b}), and
\begin{equation}
\Phi=C_B^2\frac{\xi\delta\nu_0^{(b)}}{\nu_0}\frac{\left(\frac{1}{C_B^{(b)}}-1\right)^{\frac{\xi\delta\nu_0^{(b)}}{\nu_0}-1}}{\left(C_B^{(b)}\right)^2}+
\frac{2\xi\delta\beta_{int}}{kT\nu_0}\exp{\left(\frac{2\xi\delta\beta_{int}}{kT\nu_0}C_B^{(b)}\right)},
\label{Phi}
\end{equation}
with $C_B$ given by Eq. (\ref{C_B_through_C_B^b}). Eqs. (\ref{C_B_through_C_B^b})-(\ref{Phi}), together with Eqs. (\ref{mu_A})-(\ref{mu^b}) and the conditions 
(\ref{C_A_plus_C_B}) constitute the final evolution PDE for $C_B^{(b)}$. 
Notice that the wetting/dewetting effect is fully retained in this derivation (subject to the above remark on $p_{1,2}^{(i)}$ values),
since the total equations (\ref{gamma_A}) and (\ref{gamma_B}) are used in Eq. (\ref{gamma}). 


After substitution of Eqs. (\ref{J_i}) in Eq. (\ref{h_t_eq}), the evolution PDE for the film thickness $h$ reads
\begin{equation}
h_t=\Omega M_A\frac{\partial}{\partial x}\left[\left(1-C_B\right) \frac{\partial \mu_A}{\partial x}\right]+
\Omega M_B\frac{\partial}{\partial x}\left[C_B \frac{\partial \mu_B}{\partial x}\right].
\label{h_t_eq_final}
\end{equation}
One may notice that apart from the factors $-C_B^{(b)}$ and $\left(1-C_B^{(b)}\right)$ the last two terms on the right-hand side of Eq. (\ref{C_B^b_t_eq_final})
are identical to the two terms on the right-hand side of Eq. (\ref{h_t_eq_final}).
Eqs. (\ref{h_t_eq_final}), (\ref{C_B_through_C_B^b}), (\ref{Psi}), (\ref{mu_A})-(\ref{gamma_B}) and (\ref{C_A_plus_C_B}) constitute the final PDE for $h$. After adimensionalization as described below, the closed forms of Eqs. (\ref{C_B^b_t_eq_final}) and (\ref{h_t_eq_final}) were obtained in \emph{Mathematica}.

\subsection{A model for the single-component film}

We now show the cardinal simplification of our binary alloy model to the familiar model of the single-component film with the uniform composition of the bulk and the surface. 
This is the surface diffusion model with the coupled wetting/dewetting effect. To obtain it, we set $C_A=C_A^{(b)}=C_B^{(b)}=0$ and $C_B=1$. Now the bulk diffusion and the bulk or surface solution (mixture) are not relevant. The only remaining equation of the model is Eq. (\ref{h_t_eq_final}) for the film thickness, which reduces to
the following Mullins-type equation:
\begin{equation}
h_t=\Omega M_B\frac{\partial^2}{\partial x^2}\left(-\gamma_B h_{xx}+\frac{\partial \gamma_B}{\partial h}\right),
\label{eq_h_t_1component}
\end{equation}
Here for consistency we retained the index B (it could have been omitted, since only one atomic species is now present). $\gamma_B$ is given by Eq. (\ref{gamma_B}). We perform 
the Linear Stability Analysis (LSA) of this simple model by linearizing Eq. (\ref{eq_h_t_1component}) about $h=h_0$, where $h_0$ is the film thickness immediately after the film deposition and at the start of the annealing phase.
LSA shows that the instability has a
long-wave character. The most dangerous perturbation wavelength $L_{max}$ (that maximizes the perturbation growth rate $\omega$) and its characteristic growth time scale are:
\begin{equation}
L_{max}=2\pi\left(h_0+s\right)\sqrt{\frac{\left(h_0+s\right)^2\gamma_B^{(0)}-s\left[\left(h_0+s\right)p_1^{(B)}-sp_2^{(B)}\right]}{s\left[\left(h_0+s\right)p_1^{(B)}-3sp_2^{(B)}\right]}},
\label{Lmax}
\end{equation}
\begin{equation}
T_{max}=\omega_{max}^{-1}=\frac{\left(h_0+s\right)^6\left(\left(h_0+s\right)^2\gamma_B^{(0)}-s\left[\left(h_0+s\right)p_1^{(B)}-sp_2^{(B)}\right]\right)}
{\Omega M_B s^2\left[\left(h_0+s\right)p_1^{(B)}-3sp_2^{(B)}\right]^2}.
\label{Tmax}
\end{equation}
From Eqs. (\ref{Lmax}) and (\ref{Tmax}) it follows that
$L_{max}, T_{max} >0$ at $h_0>h_{0c}\equiv s\left(3p_2^{(B)}/p_1^{(B)}-1\right)$. A film of a uniform thickness $h> h_{0c}$ is linearly unstable with respect to small thickness perturbations, and a film of a thickness $h< h_{0c}$ is linearly stable (due to the stabilizing wetting layer).

\subsection{Dimensionless model for the alloy film}

Assuming $h_0>h_{0c}$ in the binary alloy model, we use $L_{max}$ and $T_{max}$ as shown above to set the longitudinal length scale (along the substrate, or in the $x$-direction) and the time scale, respectively, in that model.
$s$ is chosen as the vertical length scale (along the $z$-axis), $g_B$ as the energy scale in the bulk, and $\gamma_B^{(0)}$ as the energy scale on the surface. After some substitutions and differentiations, the final, dimensionless problem for $C_B^{(b)}$ and $H$ is as follows:
\begin{eqnarray}
\left[H+\Delta \bar \Phi\right]\frac{\partial C_B^{(b)}}{\partial t}&=&B_M \frac{\partial}{\partial x}\left(H \frac{\partial \bar \mu^{(b)}}{\partial x}\right)-
S_{MA}C_B^{(b)} \frac{\partial}{\partial x}\left[\left(1-C_B\right) \frac{\partial \bar \mu_A}{\partial x}\right] \nonumber \\
&+&S_{MB} \left(1-C_B^{(b)}\right) \frac{\partial}{\partial x}\left[C_B \frac{\partial \bar \mu_B}{\partial x}\right]-\alpha_5\frac{\partial^4 C_B^{(b)}}{\partial x^4}, \label{C_B^b_t_eq_final_adim}
\end{eqnarray}
\begin{equation}
H_t=S_{MA}\frac{\partial}{\partial x}\left[\left(1-C_B\right)\frac{\partial \bar \mu_A}{\partial x}\right]+
S_{MB} \frac{\partial}{\partial x}\left[C_B \frac{\partial \bar \mu_B}{\partial x}\right],
\label{h_t_eq_final_adim}
\end{equation}
\begin{equation}
C_B\left(C_B^{(b)}\right)=\left[\left(\frac{1}{C_B^{(b)}}-1\right)^{\xi \alpha_1 \alpha_2}+\exp{\left(2\xi \alpha_1 \alpha_4 C_B^{(b)}\right)}+1+\bar \Psi\right]^{-1},
\label{C_B_through_C_B^b_adim}
\end{equation}
\begin{equation}
\bar \Psi= \exp{\left(\alpha_1\left[\xi\left(G_A-1-\alpha_4\right)-\frac{\Gamma_B^{(0)}}{\Delta}\left(\Gamma_A^{(0)}-1\right)\right]\right)},
\label{Psi_adim}
\end{equation}
\begin{equation}
\bar \Phi= C_B^2\xi \alpha_1 \alpha_2\frac{\left(\frac{1}{C_B^{(b)}}-1\right)^{\xi \alpha_1\alpha_2-1}}{\left(C_B^{(b)}\right)^2}+
2\xi \alpha_1\alpha_4\exp{\left(2\xi \alpha_1\alpha_4C_B^{(b)}\right)},
\label{Phi_adim}
\end{equation}
\begin{equation}
\bar \mu^{(b)}=\alpha_2 \ln{\frac{1-C_B^{(b)}}{C_B^{(b)}}}+G_A-1+\alpha_4\left(2C_B^{(b)}-1\right),\label{mu^b_adim}
\end{equation}
\begin{eqnarray}
\bar \mu_A &=& \bar g^{(b)}+\frac{\Gamma_B^{(0)}}{\Delta}C_B^{(b)}\left(\bar \gamma_A-\bar \gamma_B+\frac{\Delta}{\Gamma_B^{(0)}\alpha_1}\ln{\frac{1-C_B}{C_B}}\right)-\alpha_3\bar \gamma H_{xx}  
\nonumber \\
&+&\Gamma_B^{(0)}\frac{\partial}{\partial H}\left(\bar \gamma_A \left(1-C_B\right)+\bar \gamma_B C_B\right),\label{mu_A_adim}
\end{eqnarray}
\begin{eqnarray}
\bar \mu_B &=& \bar g^{(b)}-\frac{\Gamma_B^{(0)}}{\Delta}
\left(1-C_B^{(b)}\right)\left(\bar \gamma_A-\bar \gamma_B+\frac{\Delta}{\Gamma_B^{(0)}\alpha_1}\ln{\frac{1-C_B}{C_B}}\right)-\alpha_3\bar \gamma H_{xx}  
\nonumber \\
&+&\Gamma_B^{(0)}\frac{\partial}{\partial H}\left(\bar \gamma_A \left(1-C_B\right)+\bar \gamma_B C_B\right),\label{mu_B_adim}
\end{eqnarray}
\begin{equation}
\bar g^{(b)}=G_A\left(1-C_B^{(b)}\right)+C_B^{(b)}+\alpha_4\left(1-C_B^{(b)}\right)C_B^{(b)}+\alpha_2\left[\left(1-C_B^{(b)}\right)\ln \left(1-C_B^{(b)}\right)+C_B^{(b)}\ln C_B^{(b)}\right], 
\label{g^b_adim}
\end{equation}
\begin{equation}
\bar \gamma = \bar \gamma_A \left(1-C_B\right)+\bar \gamma_B C_B+\frac{\Delta}{\Gamma_B^{(0)}\alpha_1}\left[\left(1-C_B\right)\ln{\left(1-C_B\right)} +C_B\ln C_B\right], \label{gamma_adim}
\end{equation}
\begin{eqnarray}
\bar \gamma_A &=& \Gamma_A^{(0)}+\frac{G_2^{(A)}}{\left(H+1\right)^2}-\frac{G_1^{(A)}}{\left(H+1\right)},\label{gamma_A_adim}\\
\bar \gamma_B &=& 1+\frac{G_2^{(B)}}{\left(H+1\right)^2}-\frac{G_1^{(B)}}{\left(H+1\right)}.\label{gamma_B_adim}
\end{eqnarray}
$x$ and $t$ now stand for the dimensionless space and time variables, $H$ is the dimensionless film thickness defined as $H=h/s$, and 
other dimensionless quantities are marked with an overbar.
Notice that the linear term $-\alpha_5\partial^4 C_B^{(b)}/\partial x^4$ was added to Eq. (\ref{C_B^b_t_eq_final_adim}), where $\alpha_5$ is the dimensionless Cahn-Hilliard gradient energy
coefficient.
The sole role of this term is to suppress growth of the short-wavelength modes, preventing the unphysical instability on the finest spatial scale \cite{LK}. This term does not affect
the emergence and the nonlinear growth/saturation of the physical instabilities, which result in changes of the concentrations and the film thickness. All values of the
physical and dimensionless parameters are collected in Tables \ref{table1} and \ref{table2}.

\emph{To summarize the above-presented derivation and the final system}, PDE (\ref{C_B^b_t_eq_final_adim}) is effectively the diffusion equation for the bulk concentration 
$C_B^{(b)}$ of B atoms in the film domain bounded above by the surface whose shape $H$ is time-dependent. Using integration across the film and mass conservation, this localized equation is derived by accounting for changes of B atoms surface concentration due to surface diffusion, the flux from the bulk, and the surface displacement (Eq. (\ref{C_B_eq})). Eq. (\ref{C_B^b_t_eq_final_adim}) 
is linked to the \emph{quasistatic} relation (\ref{C_B_through_C_B^b_adim}) between the bulk and surface concentrations that expresses segregation of one component to the surface.
The second PDE of the system, Eq. (\ref{h_t_eq_final_adim}), describes the evolution of the surface shape by surface diffusion. This equation is also linked to the ``segregation" equation
(\ref{C_B_through_C_B^b_adim}). Substitution of the segregation equation in both PDEs, followed by the substitution of Eqs. (\ref{Psi_adim})-(\ref{gamma_B_adim}) yield the final, closed dimensionless forms, which are coupled through $C_B^{(b)}$ and $H$ that appear in both
equations. The dynamics of the bulk concentration and the surface shape is therefore coupled. Also, once $C_B^{(b)}$ has been computed from Eq. (\ref{C_B^b_t_eq_final_adim}), the dynamics of the surface concentration $C_B$ is obtained from the segregation equation simply by substituting $C_B^{(b)}$. Wetting/dewetting potentials, the bulk diffusion and the phase separation, and other thermodynamic effects 
(such as the typically unequal bulk and surface energies $\bar g^{(b)}$ and $\bar \gamma$) and the kinetic effects (such as the often unequal surface mobilities of A and B atoms) are accounted for in the derivation.

\section{Remarks on the bulk phase separation and the surface segregation}
\label{Consider}

\begin{figure}[H]
\vspace{-0.4cm}
\centering
\includegraphics[width=5.5in]{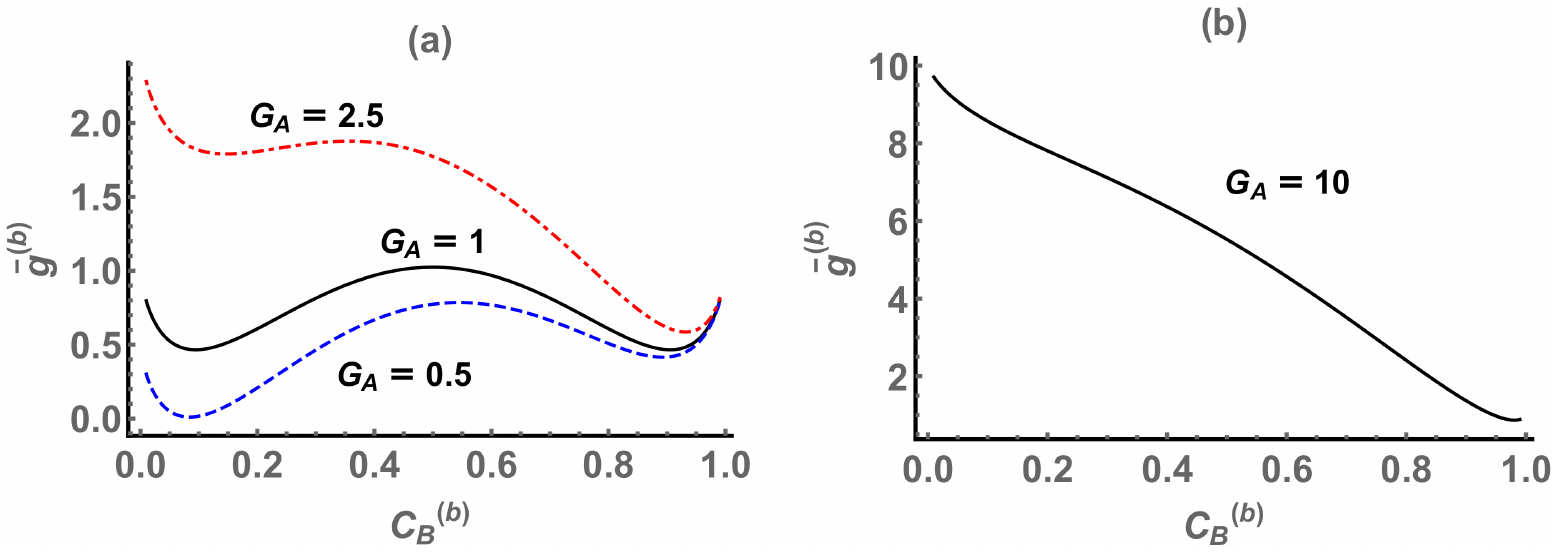}
\vspace{-0.15cm}
\caption{(Color online.) The dimensionless bulk energy. 
}
\label{Fig1}
\end{figure}

The dimensionless bulk energy, Eq. (\ref{g^b_adim}) is plotted in Fig. \ref{Fig1} for the parameters from Table \ref{table2}. For $0.5 \le G_A \le 2.5$ the double-well form of 
this curve in panel (a) suggests the bulk phase separation into the regions of large and small 
concentration of B atoms. According to panel (b) for $G_A>2.5$ there is no phase separation. Since the nonlinear evolution PDEs (\ref{C_B^b_t_eq_final_adim}) and (\ref{h_t_eq_final_adim}) are heavily coupled, the process of phase separation affects (and in turn, is influenced by) the dewetting of the film and the formation of the particles composed of a mixture of A and B atoms. Of course, even when the phase separation is not present, the redistribution of the bulk composition still takes place through bulk diffusion.
\vspace{-0.75cm}
\begin{figure}[H]
\centering
\includegraphics[width=4.75in]{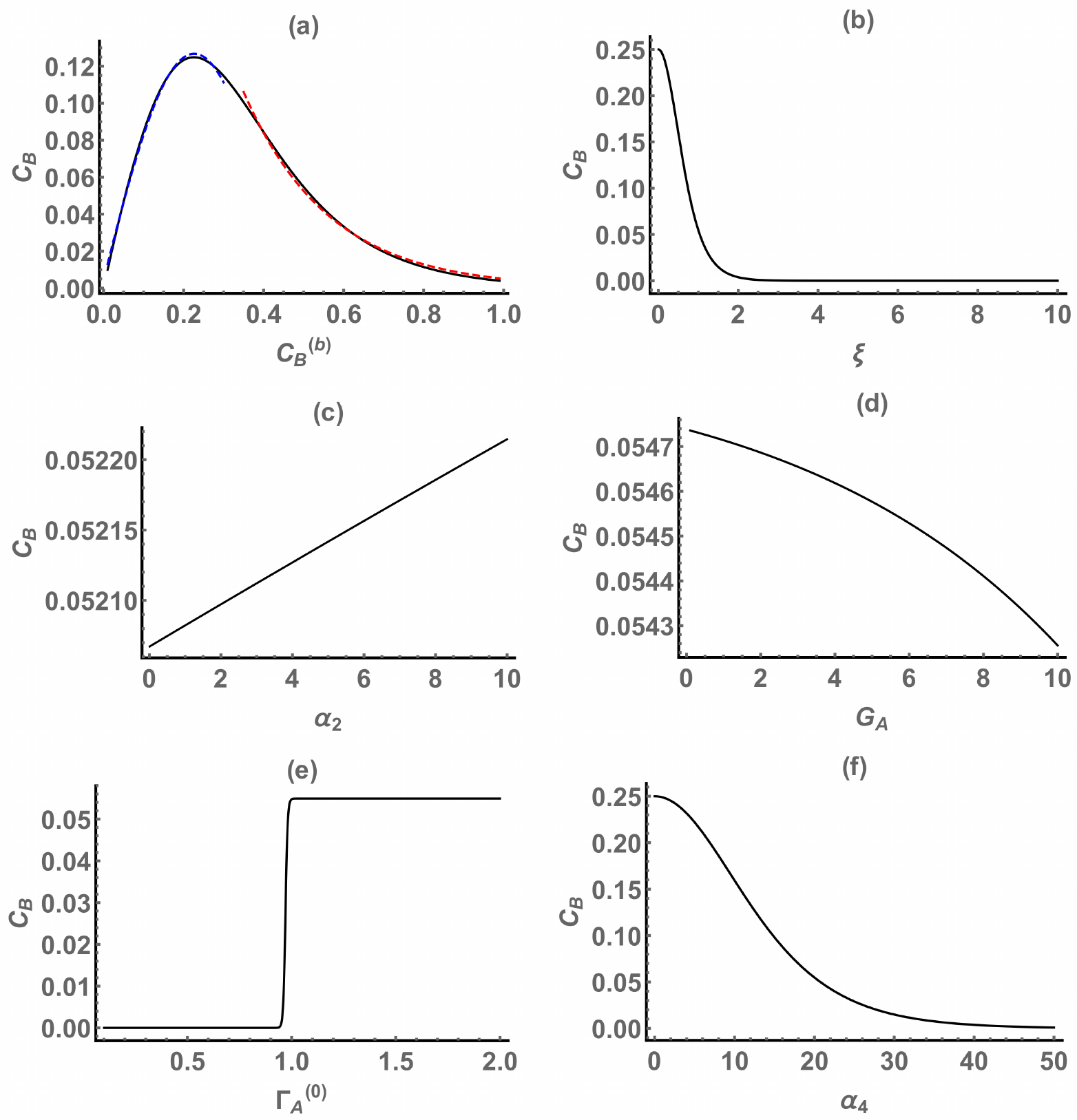}
\vspace{-0.15cm}
\caption{(Color online.) $C_B$ vs. $C_B^{(b)}$ and the parameters entering Eq. (\ref{C_B_through_C_B^b_adim}). In (a), the dashed lines show the fits:
$C_B=0.1276 \sin{7.0235 C_B^{(b)}},\ 0<C_B^{(b)}\le 0.3$; $C_B=0.5477 \exp{\left(-4.689 C_B^{(b)}\right)},\ 0.35\le C_B^{(b)}< 1$.  $C_B^{(b)}=0.5$ in (b)-(f).
}
\label{Fig9}
\end{figure}
\begin{figure}[H]
\centering
\includegraphics[width=7.0in]{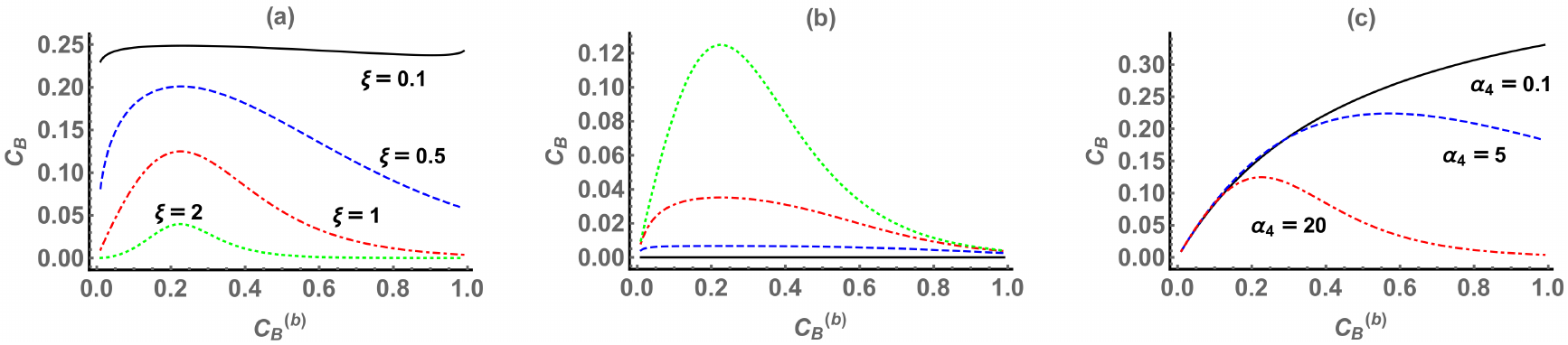}
\vspace{-0.15cm}
\caption{(Color online.) $C_B$ vs. $C_B^{(b)}$ at $\xi$, $\Gamma_A^{(0)}$ and $\alpha_4$ varied. In (b) the solid, dashed, dashed-dotted, and
dotted lines correspond to $\Gamma_A^{(0)}=0.9,\ 0.96,\ 0.97,\ 1$, respectively.
}
\label{Fig91}
\end{figure}
\begin{table}[!ht]
\centering
{\scriptsize 
\begin{tabular}
{|c|c|}

\hline
				 
			\rule[-2mm]{0mm}{6mm} \textbf{Physical parameter ($i=A,\ B$)}	 & \textbf{Typical value}  \\
			\hline
                        \hline
			\rule[-2mm]{0mm}{6mm} $s$ & $0.5\times 10^{-7}$ cm (2 ML) \cite{NMPBA}  \\
                        \hline
			\rule[-2mm]{0mm}{6mm} $h_0$ & $1.2\times 10^{-6}$ cm (12 nm) \cite{AKR}  \\
			\hline
                        \rule[-2mm]{0mm}{6mm} $\delta$ & $10^{-8}$ cm \cite{ZVD}  \\
			\hline
				\rule[-2mm]{0mm}{6mm} $\nu_i^{(b)}=\rho N_{Av}/M$ & $5\times 10^{21}$ cm$^{-3}$  \cite{ZVD}  \\
                        \hline
				\rule[-2mm]{0mm}{6mm} $\nu_0^{(b)}=\nu_A^{(b)}+\nu_B^{(b)}$ & $10^{22}$ cm$^{-3}$  \cite{ZVD}  \\
			\hline
				\rule[-2mm]{0mm}{6mm} $\nu_i$ & $0.5\times 10^{14}$ cm$^{-2}$ \cite{ZVD} \\
                        \hline
				\rule[-2mm]{0mm}{6mm} $\nu_0=\nu_A+\nu_B$ & $10^{14}$ cm$^{-2}$ \cite{ZVD} \\
                        \hline
			\rule[-2mm]{0mm}{6mm} $\Omega=1/\nu_0^{(b)}$ & $10^{-22}$ cm$^3$ \\
			\hline
			   \rule[-2mm]{0mm}{6mm} $D_i$ & $10^{-7}$ cm$^2/$s  \cite{AKR} 
                                                                                         \\
                          \hline
                         \rule[-2mm]{0mm}{6mm} $D_B^{(b)}$ & $10^{-11}$ cm$^2/$s \cite{AKR} \\
                          \hline
			\rule[-2mm]{0mm}{6mm} $T$ & 923 K (650 $^{\circ}$C) \cite{AKR} \\
			\hline
			    \rule[-2mm]{0mm}{6mm} $\gamma_B^{(0)}$ & $2.5\times 10^3$  erg$/$cm$^2$ \cite{SR}  \\
			\hline
				\rule[-2mm]{0mm}{6mm} $g_B$ & $1.8\times 10^8$ erg$/$cm$^3$  \cite{NIST} \\
                        \hline
				\rule[-2mm]{0mm}{6mm} $p_1^{(i)}$ & $2.5$  erg$/$cm$^2$ \cite{SZ} \\ 
                        \hline
				\rule[-2mm]{0mm}{6mm} $p_2^{(i)}$ & $2.5$  erg$/$cm$^2$ \cite{SZ} \\ 
			\hline
			    \rule[-2mm]{0mm}{6mm} $\beta_{int}$ & $3.6\times 10^9$ erg$/$cm$^3$ \cite{ZVD}  \\
                        \hline
			    \rule[-2mm]{0mm}{6mm} $L_{max}$ (based on $h_0=10$ nm) & $10^{-3}$ cm  \\
                        \hline
			    \rule[-2mm]{0mm}{6mm} $T_{max}$ (based on $h_0=10$ nm) & $7.5\times 10^5$ s  \\ 
			\hline

\end{tabular}}
\caption[\quad Physical parameters]{Physical parameters (see the definitions in Sec. \ref{Formulation}). When possible, the cited values are for Fe.} 
\label{table1}
\end{table}
\begin{table}[!ht]
\centering
{\scriptsize 
\begin{tabular}
{|c|c|c|c|}

\hline
				 
			\rule[-2mm]{0mm}{6mm} \textbf{Dimensionless parameter ($i=A,\ B$)} & \textbf{Typical value}  & \textbf{Description} & \textbf{Fixed or Variable} \\
			\hline
                        
                        \hline
			\rule[-2mm]{0mm}{6mm} $\Delta=\delta/s$ & $0.2$ & Thickness of the surface layer & Fixed \\

                        \hline
			\rule[-2mm]{0mm}{6mm} $H_0=h_0/s$ & $24$ & Nominal (initial) film height & Fixed \\

                        \hline
			\rule[-2mm]{0mm}{6mm} $C_B^{(b)(0)}$ & $0.5$ & Nominal (initial) concentration of B atoms & Fixed \\

                        \hline
			\rule[-2mm]{0mm}{6mm} $B_M=\Omega M_B^{(b)} g_B T_{max}/L_{max}^2$ & $0.98$ & Bulk mobility of B atoms  & Fixed \\

			\hline
				\rule[-2mm]{0mm}{6mm} $S_{Mi}=\Omega M_i g_B T_{max}/s L_{max}^2$ & $985.25$ & Surface mobilities of A,B atoms & Variable \\
			\hline
				\rule[-2mm]{0mm}{6mm} $G_A=g_A/g_B$ & $1$ & Ratio of bulk energies  & Variable \\
			\hline
			   \rule[-2mm]{0mm}{6mm} $\Gamma_A^{(0)}=\gamma_A^{(0)}/\gamma_B^{(0)}$ & $1$ & Ratio of surface energies & Variable \\
			\hline
			\rule[-2mm]{0mm}{6mm} $\Gamma_B^{(0)}=\gamma_B^{(0)}/s g_B$ & $277.78$ & Ratio of B atoms surface energy to bulk energy & Fixed \\
			\hline
				\rule[-2mm]{0mm}{6mm} $G_1^{(i)}=p_1^{(i)}/\gamma_B^{(0)}$ & $10^{-3}$ & Ratios of wetting energies to B atoms surface energy & Fixed \\
			\hline
			    \rule[-2mm]{0mm}{6mm} $G_2^{(i)}=p_2^{(i)}/\gamma_B^{(0)}$ & $10^{-3}$ & Ratios of wetting energies to B atoms surface energy & Fixed \\
			
			\hline
			   \rule[-2mm]{0mm}{6mm} $\xi$ & $1$ & Ratio of contributions to the flux (Eq. (\ref{from_F_B})) & Variable \\
			\hline
				\rule[-2mm]{0mm}{6mm} $\alpha_1=\delta g_B/k T \nu_0$ & $0.14$ & Ratio of B atoms bulk energy to surface entropy & Fixed \\
			\hline
			    \rule[-2mm]{0mm}{6mm} $\alpha_2=k T\nu_0^{(b)}/g_B$ & 7.18 & Ratio of bulk entropy to B atoms bulk energy & Fixed \\
			\hline
				\rule[-2mm]{0mm}{6mm} $\alpha_3=(s/L_{max})^2 \Gamma_B^{(0)}$ & $6.51\times 10^{-7}$ & Scaled $\Gamma_B^{(0)}$  & Fixed \\
			\hline
			    \rule[-2mm]{0mm}{6mm} $\alpha_4=\beta_{int}/g_B$ & 20  & Ratio of bulk enthalpy to B atoms bulk energy  & Fixed \\
                        \hline
				\rule[-2mm]{0mm}{6mm} $\alpha_5=\Omega M_B^{(b)} \kappa T_{max}\left(\nu_B^{(b)} V\right)/s L_{max}^4$ \cite{LJS} & $10.27$ & Short-wavelength instability cut-off parameter & Fixed \\
                        
			\hline				

\end{tabular}}
\caption[\quad Dimensionless parameters]{Dimensionless parameters. Typical values are based on values in Table I. In the definitions of $B_M$ and $S_{Mi}$, $M_i=\Omega \nu_i D_i/kT$ and $M_B^{(b)}=\Omega \nu_B^{(b)} D_B^{(b)}/kT$ are the dimensional surface and bulk mobilities, respectively. 
In the definition of $\alpha_5$, $V=h_0\times \left(100h_0\right)^2$ is the characteristic volume of the system, and $\kappa=10^{-12}$ erg \cite{LK}.
} 
\label{table2}
\end{table}

In Fig. \ref{Fig9} the dimensionless concentration $C_B$ from Eq. (\ref{C_B_through_C_B^b_adim}) is plotted vs. $C_B^{(b)}$ and the various dimensionless parameters that enter this equation.
The spatio-temporal variation of $C_B$ and $C_B^{(b)}$ is ignored. 
In each panel one parameter is changing along the horizontal axis, while all other parameters are at their values stated in Table \ref{table2}. It can be seen that $C_B$
is small, that is, A atoms segregate from the bulk to the surface. 
The largest variations of
$C_B$ (up to 25\%) occur when $C_B^{(b)}$, $\xi$, and $\alpha_4$ are varied. Using the plots in this figure, the final mean value of $C_B$ in the dynamical computation of dewetting 
(Sec. \ref{Comput}) can be fairly accurately predicted for a particular set of the parameters values. For instance, Fig. \ref{Fig5} of Sec. \ref{Comput} was computed at $\{$ $C_B^{(b)}=0.5$, 
$\xi=1$, $\alpha_2=7.2$, $G_A=1$, $\Gamma_A^{(0)}=1$, $\alpha_4=20$ $\}$, and the mean value of $C_B$ in Fig. \ref{Fig5}(c) 
is around 0.006. Now notice that the same $C_B$ value is seen in Fig. \ref{Fig9}(a) at $C_B^{(b)}=0.5$, in Fig. \ref{Fig9}(b) at $\xi=1$, in Fig. \ref{Fig9}(c) at $\alpha_2=7.2$, in Fig. \ref{Fig9}(d) at $G_A=1$, in Fig. \ref{Fig9}(e) at $\Gamma_A^{(0)}=1$,
and in Fig. \ref{Fig9}(f) at $\alpha_4=20$.  

The interesting dependence of $C_B$ on $C_B^{(b)}$ in Fig. \ref{Fig9}(a) raises the question of how this dependence changes when one of the key parameters is varied.
From the other panels in Fig. \ref{Fig9} it can be seen that $\xi$, $\Gamma_A^{(0)}$ and $\alpha_4$ have large impacts on $C_B$, thus in Fig. \ref{Fig91} $C_B$
is plotted vs. $C_B^{(b)}$ at one of these parameters varying and the other two parameters and all other parameters fixed to the Table \ref{table2} values. Fig. \ref{Fig91} shows 
that the characteristic shape of the $C_B\left(C_B^{(b)}\right)$ function (the growth followed by the decay) is persistent. In panel (a) the maximum of
$C_B$ is attained at roughly the same $C_B^{(b)}$ value ($\sim 0.22$), with $C_B$ increasing faster and decaying slower when $\xi$ is decreased;  in panel (b) the pronounced maximum
of the curve is developing with the increase of $\Gamma_A^{(0)}$, and in panel (c) the monotonic growth is replaced by the growth-decay when the dimensionless bulk enthalpy is large.
Notice also that $C_B$ indirectly depends (through $C_B^{(b)}$)
on the kinetic parameters $B_M$, $S_{Mi}$, as well as on the wetting/dewetting potentials strengths $G_1^{(i)}$, $G_2^{(i)}$.

Further, Ref. \cite{AR} cites the expression for the saturation time, $t_{sat}$, needed to achieve the equilibrium surface concentration in a system with a strong segregation tendency:
$t_{sat}=\pi s^2 d^2/4D$, where $s\sim 4$ and $d\sim 0.7$ nm are the surface enrichment factor and the width of the surface segregation layer, and $D$ is the bulk diffusivity of 
the segregating species.
(Notice that $d$ and $\delta$ are not the same widths.) Using value of the bulk diffusivity from Table \ref{table1} gives $t_{sat}=0.006$ s, which is significantly shorter than
the evolution times needed to form the particles in the simulations (Sec. \ref{Comput}). Same conclusion is in Ref. \cite{AR}, where even shorter saturation time is noted. Thus one has to keep in mind that the profiles of $C_B$ at various times in 
Figures \ref{Fig5}-\ref{Fig7} show how the \emph{equilibrium} surface concentration evolves in response to the changes of the bulk composition and surface morphology.

\section{Morphological and compositional stability of the binary alloy film}
\label{ResultsLSA}

In this section we briefly discuss the results of LSA. The linearization is performed about the
$\left(H_0,C_B^{(b)(0)}\right)$ pair from Table \ref{table2}. Our focus is on studying how the variations of  
$G_A$, $\Gamma_A^{(0)}$, and $\xi$ affect the stability of the planar surface morphology and the uniform composition of the alloy film.

\begin{figure}[h]
\centering
\includegraphics[width=6.75in]{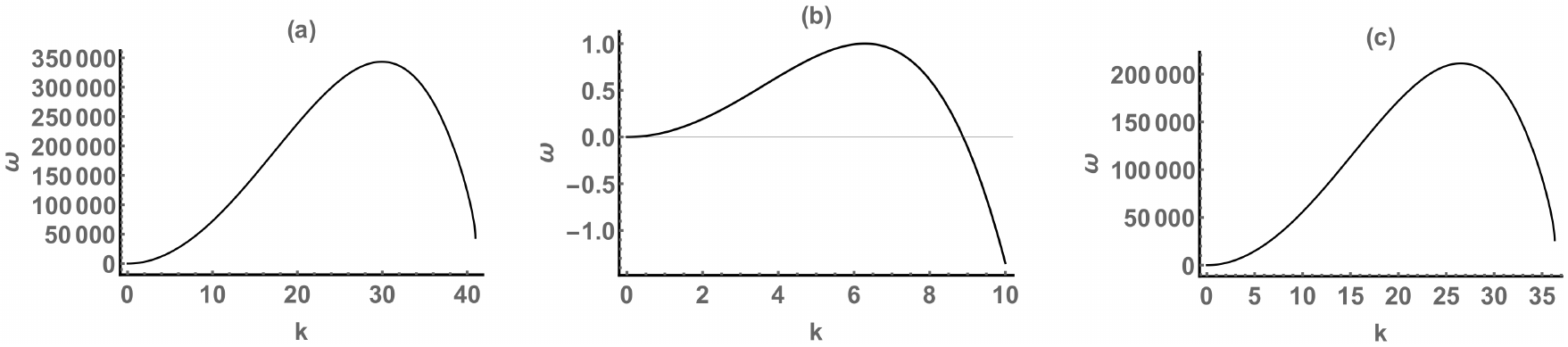}
\vspace{-0.15cm}
\caption{Typical perturbation growth rate $\omega(k)$ for (a) the alloy film with the bulk phase separation $(G_A=1)$, (b) the single-component film, and (c) the alloy film without the bulk phase separation
$(G_A=10)$ (ref. Fig. \ref{Fig1}). $\Gamma_A^{(0)}=\xi=1$.}
\label{Fig2}
\end{figure}

In Fig. \ref{Fig2} the typical long wave dispersion curve $\omega(k)$ for the alloy film is compared to the one for the single-component film; the latter is represented by the 
adimensionalized Eq. (\ref{eq_h_t_1component}). Here $\omega$ is the growth rate, and $k$ is the wavenumber of the normal perturbation mode in the direction along the substrate. 
Notice that the ``single-component" curve does not change when $G_A$, $\Gamma_A^{(0)}$, and $\xi$ are varied, since these parameters do not enter
the single-component model. It can be seen that the most dangerous wavelength of the instability $\lambda_{max}=2\pi/k_{max}$ (where $k_{max}$ is the wavenumber at the curve's maximum)
is smaller for the alloy film, and the corresponding perturbation growth rate $\omega_{max}$ is much larger than the one for the single-component film.
The nonlinear development of the instability (computed in Sec. \ref{Comput}) 
results in the dewetted film, that is, the particles of the average size $\lambda_{max}$, located on top of the wetting layer, and composed of the mixture of A and B atoms. This structure may undergo slow coarsening.
Notice also that the case of the alloy film without the bulk phase separation (Fig. \ref{Fig2}(c))
is intermediate between the single-component film and the phase-separating alloy film. 

\begin{figure}[h]
\centering
\includegraphics[width=6.75in]{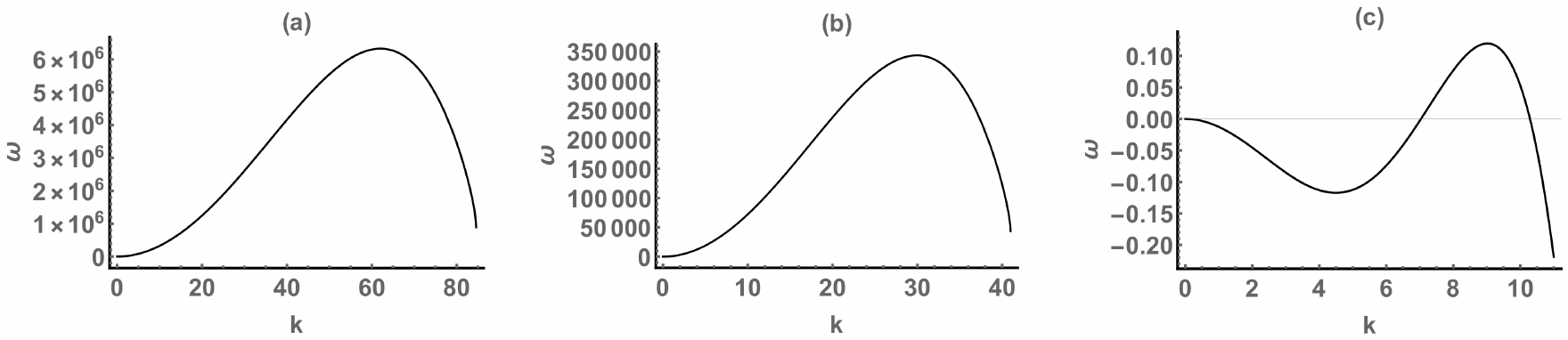}
\vspace{-0.15cm}
\caption{Perturbation growth rate for (a) $\Gamma_A^{(0)}=1.1$, (b) $\Gamma_A^{(0)}=1$, and (c) $\Gamma_A^{(0)}=0.97$.
}
\label{Fig3}
\end{figure}

Table \ref{table3} summarizes LSA results for $\xi$ variation. As $\xi$ decreases, $\lambda_{max}$ monotonically increases, and $\omega_{max}$ monotonically decreases. Table \ref{table4} summarizes LSA results for $G_A$ variation. As $G_A$ increases, $\lambda_{max}$ monotonically increases, and $\omega_{max}$ monotonically decreases. These changes are very minor compared to the effects of $\xi$ variation. The most drastic, qualitative impacts on the instability are observed when $\Gamma_A^{(0)}$ is varied.
They are shown in Fig. \ref{Fig3}. When $\Gamma_A^{(0)}$ decreases, first $\lambda_{max}$ increases and $\omega_{max}$ decreases 
(Fig. \ref{Fig3}(a,b)), and then at $\Gamma_A^{(0)}=0.97$ the instability changes from the long wave to the short wave.
The short wave character of the instability persists when $\Gamma_A^{(0)}$ is further decreased to $\sim 0.5$. In Table \ref{table5} the summary of LSA results is provided for the case
of $\Gamma_A^{(0)}$ variation. Notice the order of magnitude change of $\lambda_{max}$ when $\Gamma_A^{(0)}$ approaches one.
If the instability is long wave, then at the instability threshold there emerge all destabilizing modes whose wavelengths are larger than $2\pi/k_c$, where $k_c$ is the positive solution of the equation 
$\omega(k)=0$. If the instability is short wave, then the spectrum of the unstable wavelengths is finite, $2\pi/k_{c2}<\lambda<2\pi/k_{c1}$, where $k_{c1},\ k_{c2}\ (k_{c2}>k_{c1})$ are two positive solutions of the equation 
$\omega(k)=0$, see Fig. \ref{Fig3}(c). The long wave instability is of the spinodal decomposition type, while the short wave one is of the Turing type. In the latter situation, after the particles were formed through film dewetting, further evolution may cease due to absence of coarsening (Ostwald ripening)  \cite{GolovinPRB2004}. 

The dimensionless values of 
$\lambda_{max}$ in Tables \ref{table3}-\ref{table5} translate into the physical values in the range 0.5 $\mu$m - 8.1 $\mu$m. This is the particle size before the coarsening sets in, and the order of magnitude of these values agrees with the experiment \cite{AKR1}. These values are 5-10 times smaller than for the single-component film, as follows from the $L_{max}$ value in Table \ref{table1} and from Fig. \ref{Fig2}. 
\begin{table}[!ht]
\centering
{\scriptsize 
\begin{tabular}
{|c|c|c|}

\hline
				 
			\rule[-2mm]{0mm}{6mm} $\xi$ & $\lambda_{max}$  & $\omega_{max}$ \\
			\hline
                        
                        \hline
			\rule[-2mm]{0mm}{6mm} 10 & 0.064 & $3.65\times 10^7$ \\

                        \hline
			\rule[-2mm]{0mm}{6mm} 5 & 0.09 & $9.56\times 10^6$ \\

                        \hline
			\rule[-2mm]{0mm}{6mm} 2 & 0.143 & $1.58\times 10^6$ \\ 

                        \hline
			\rule[-2mm]{0mm}{6mm} 1 & 0.21 & $3.434\times 10^5$ \\

			\hline
				\rule[-2mm]{0mm}{6mm} 0.5 & 0.32 & $6.1\times 10^4$ \\
			\hline
				\rule[-2mm]{0mm}{6mm} 0.1 & 0.53 & $8.5\times 10^3$ \\
                        
			\hline				

\end{tabular}}
\caption[\quad LSA results, part I]{LSA results for varying $\xi$. $G_A=\Gamma_A^{(0)}=1$.} 
\label{table3}
\end{table}
\begin{table}[!ht]
\centering
{\scriptsize 
\begin{tabular}
{|c|c|c|}

\hline
				 
			\rule[-2mm]{0mm}{6mm} $G_A$ & $\lambda_{max}$  & $\omega_{max}$ \\
			\hline

                        \hline

			\rule[-2mm]{0mm}{6mm} 0.5 & 0.208 & $3.516\times 10^5$ \\

                        \hline
			\rule[-2mm]{0mm}{6mm} 1 & 0.21 & $3.434\times 10^5$ \\

                        \hline
			\rule[-2mm]{0mm}{6mm} 1.5 & 0.211 & $3.352\times 10^5$ \\ 

                        \hline
			\rule[-2mm]{0mm}{6mm} 2 & 0.212 & $3.272\times 10^5$ \\

			\hline
				\rule[-2mm]{0mm}{6mm} 2.5 & 0.214 & $3.193\times 10^5$ \\
			                        
			\hline				

\end{tabular}}
\caption[\quad LSA results, part II]{LSA results for varying $G_A$. $\xi=\Gamma_A^{(0)}=1$.} 
\label{table4}
\end{table}
\begin{table}[!ht]
\centering
{\scriptsize 
\begin{tabular}
{|c|c|c|}

\hline
				 
			\rule[-2mm]{0mm}{6mm} $\Gamma_A^{(0)}$ & $\lambda_{max}$  & $\omega_{max}$ \\
			\hline

                        \hline

			\rule[-2mm]{0mm}{6mm} 0.5 & 0.69 & 0.006 \\

                        \hline
			\rule[-2mm]{0mm}{6mm} 0.7 & 0.76 & 0.00005 \\

                        \hline
			\rule[-2mm]{0mm}{6mm} 0.9 & 0.81 & 0.0015 \\ 

                        \hline
			\rule[-2mm]{0mm}{6mm} 1 & 0.21 & $3.434\times 10^5$ \\

			\hline
				\rule[-2mm]{0mm}{6mm} 1.2 & 0.076 & $1.974\times 10^7$ \\

                        \hline
				\rule[-2mm]{0mm}{6mm} 1.4 & 0.056 & $6.886\times 10^7$ \\
			                        
			\hline				

\end{tabular}}
\caption[\quad LSA results, part III]{LSA results for varying $\Gamma_A^{(0)}$. $\xi=G_A=1$.} 
\label{table5}
\end{table}

\section{Computations of the morphological and compositional evolution of the alloy film}
\label{Comput}

For computations, we take the horizontal dimension of the surface $L=20\lambda_{max}$, where $\lambda_{max}$ is computed from LSA for each set of parameters.
The boundary conditions for $H(x,t),\ C_B^{(b)}(x,t)$ are periodic at $x=0,\ L$.

First, we present computed results for the long wave instability case shown in Figures \ref{Fig2}(a) and \ref{Fig3}(b) (the same figure). Our goal is to illustrate how various engineered initial conditions (mainly, the initial inhomogeneous distribution of the bulk concentration) 
influence film dewetting, the formation of particles, and the redistribution of the alloy components inside the particles and on their surface. 
This focus is because the initial inhomogeneous distributions of the bulk concentration are fairly easy to engineer in the experiment by 
precision alloying. It should be kept in mind that the particles that formed after the film dewetted  
may undergo a slow coarsening by the mass exchange through the wetting layer. This coarsening we do not attempt to characterize. 
Because the time scale is known, in the experiment it would be possible to ``freeze" the evolution at any stage by rapidly cooling the system down, and thus one may, in principle, obtain the particles distribution as desired.
We remark that the film volume is conserved in all computations; the maximum volume change does not exceed 0.1\%.

\begin{figure}[h]
\centering
\includegraphics[width=6in]{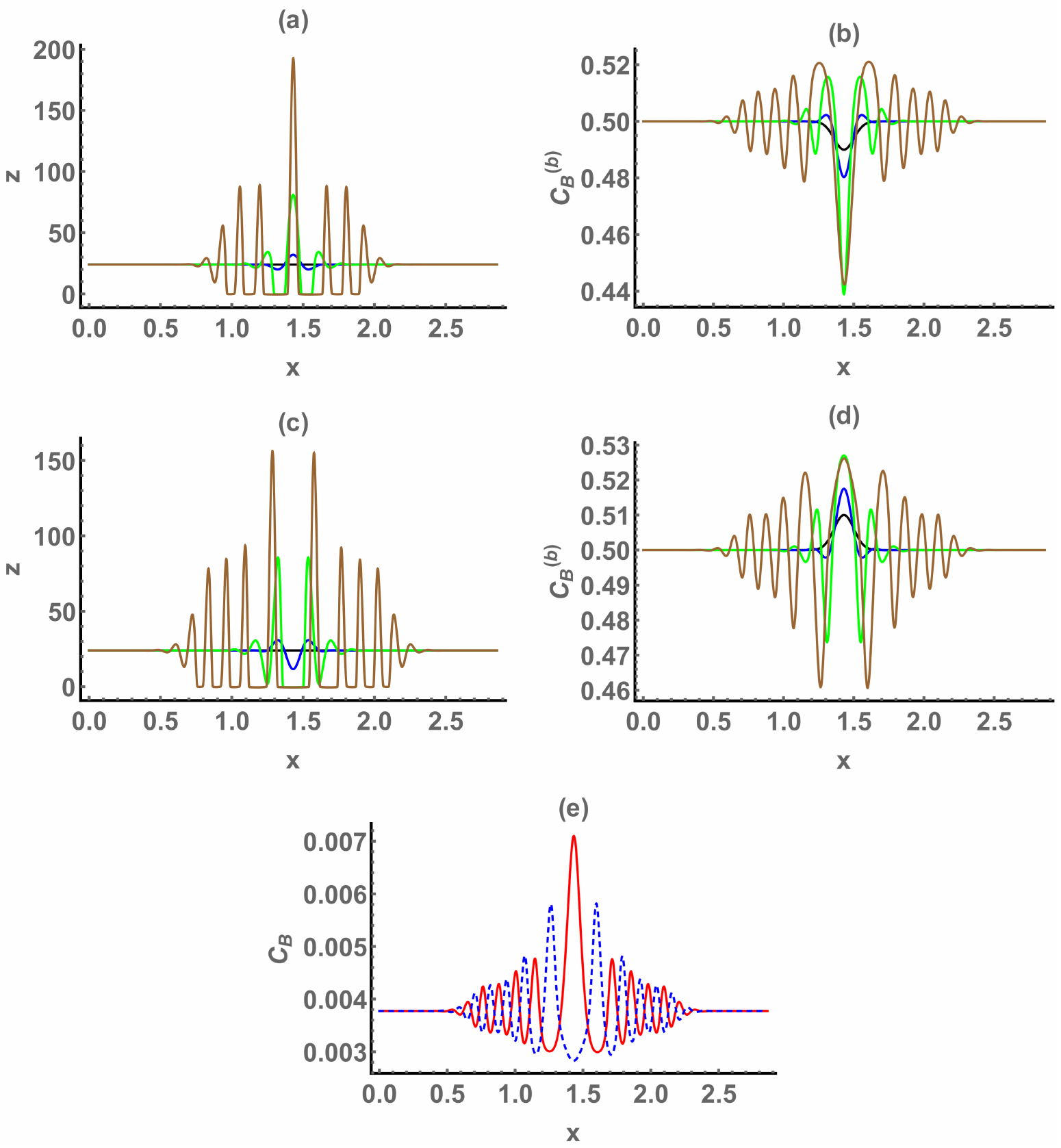}
\vspace{-0.15cm}
\caption{(Color online.) Evolution of the small Gaussian spike of the bulk concentration (b,d) and the corresponding film morphology (a,c) (initially the planar surface).
The parameters in the Gaussian are $W=0.1$, $L=2.86$ and (b) $\eta=-0.01$, (d) $\eta=0.01$. The final time $t_f=1.6\times 10^{-4}$ (2 min). 
The final profiles of the surface concentration for these cases are shown in (e),
with the solid (dashed) line corresponding to $\eta=-0.01$ ($\eta=0.01$). 
}
\label{Fig4}
\end{figure}
\begin{figure}[H]
\centering
\includegraphics[width=6.75in]{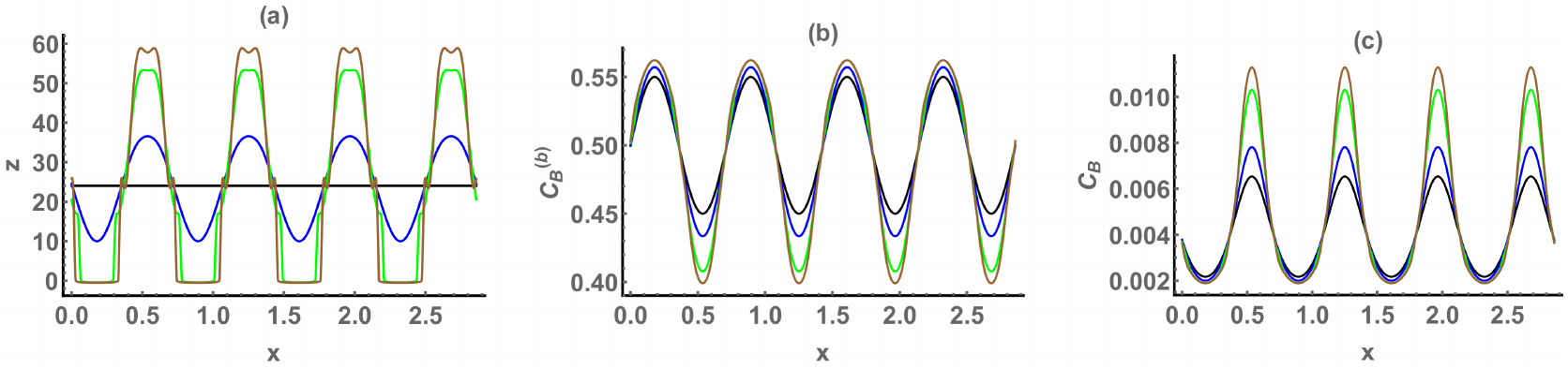}
\vspace{-0.15cm}
\caption{(Color online.) (b) Evolution of the periodic profile of the bulk concentration, (a) the corresponding film morphology (initially the planar surface) and (c)
the profiles of the surface concentration.
The final time $t_f=5\times 10^{-5}$ (38 s). 
 }
\label{Fig5}
\end{figure}

Figures \ref{Fig4}(b,d) show the evolution of the bulk concentration starting from the initial condition $H(x,0)=H_0=24$, 
$C_B^{(b)}(x,0)=0.5+\eta\exp{\left[-(x-L/2)^2/W^2\right]}$ (the Gaussian-shaped curve) with $\eta=-0.01$ $(\eta=0.01)$, respectively; this is the deviation of 1\% from the 
uniform bulk composition (50\%(A)$/$50\%(B)). The corresponding formation of the particles is shown in Figures \ref{Fig4}(a,c). In the top row ($\eta<0$) the central particle is formed precisely at the location along the substrate where the initial $C_B^{(b)}$ is depressed,
and as the evolution proceeds this growing particle is gradually enriched by the A phase, until the equilibrium is achieved between the particle size and the composition. Locations of the smaller particles to the left or right of the central particle also correlate with the minima of $C_B^{(b)}$, and thus they also are rich in the A phase, while the wetting layer 
between the particles is rich in the B phase. The change of $C_B^{(b)}$ from a minimum to a maximum relative to the initial value 1/2 is 16\%. The evolution in the middle row (Figures \ref{Fig4}(c,d), $\eta>0$) bears many similarities to the one in the top row, but
at the same time we notice the important differences.  Here the particles also are rich in the A phase, but their spatial arrangement is not the same. The central particle is absent
altogether, and the satellite particles have been formed at the location where in Fig. \ref{Fig4}(a) there is the wetting layer.  Thus \emph{the first important observation} is that by precision engineering the initial concentration profile, it should be possible to 
guide not only the particles composition, but also their spatial localization. In Fig. \ref{Fig4}(e) the final surface concentrations corresponding to the two discussed cases are compared. In either case A atoms segregated from the bulk to the surface, as is expected from Fig. \ref{Fig9}, and the surface composition is a nearly homogeneous A phase. The shapes of the final surface concentration profiles 
basically are the reflections of the bulk concentration profiles in the horizontal line (the mirror images). Thus 
\emph{the second important observation} is that the particles are core-shell, where the shell is the nanometric segregated layer of the A phase,
and the core is the alloy that is modestly rich in the A phase. This matches the experiment \cite{AKR,AKR1,AR}.
As in the experiment, here the computed surface segregation of the A phase is the thermodynamic effect caused by the unequal surface and bulk energies \cite{Tersoff}; 
notice that the kinetic effects are absent, since the surface mobilities are taken equal. Indeed, after the wetting effect is eliminated for simplicity by taking $\bar \gamma_A=\Gamma_A^{(0)}$, $\bar \gamma_B=1$, equations 
(\ref{g^b_adim}) and (\ref{gamma_adim}) give at the nominal concentration $C_B^{(b)}=0.5$ and the parameters from Table \ref{table2}: $\bar \gamma=0.999$, $\bar g_B^{(b)}=1.024$. 
Also notice that although the computation is done with the bulk phase separation activated, turning it off (by taking the ratio $G_A$ of the components' bulk energies 
outside of the interval $0.5\le G_A\le 2.5$, or by taking the zero value for the dimensionless alloy entropy $\alpha_2$ while keeping $G_A$ in the interval) does not have a pronounced effect on the particles morphologies or
compositions, but the characteristic evolution time scale increases. Say, taking $G_A=10$ instead of $G_A=1$ in the simulation shown in Fig. \ref{Fig5} resulted in the morphologies and compositions similar to those in this Figure, but the final time increased threefold. This is consistent with LSA, see the discussion of Fig. \ref{Fig2}.
The simulation without the bulk phase separation also points out the significance of the bulk diffusion; the same conclusion is reached in the experiment \cite{AKR1} (page 5140). The bulk diffusion was also considered in Ref. \cite{RS} within the linear dynamical framework that assumes small deviations of the bulk compositions and the surface profile from the initial values. These authors found that
the bulk diffusion is effective in suppressing decomposition induced by the difference in surface diffusivities of the alloy components, thereby reducing the effect of the kinetic 
surface segregation.

In Figures \ref{Fig5}(a-c) some of the features of Fig. \ref{Fig4} can be observed; for example, the particles form at the locations where $C_B^{(b)}$ attains a minimum.

\begin{figure}[H]
\centering
\includegraphics[width=5.5in]{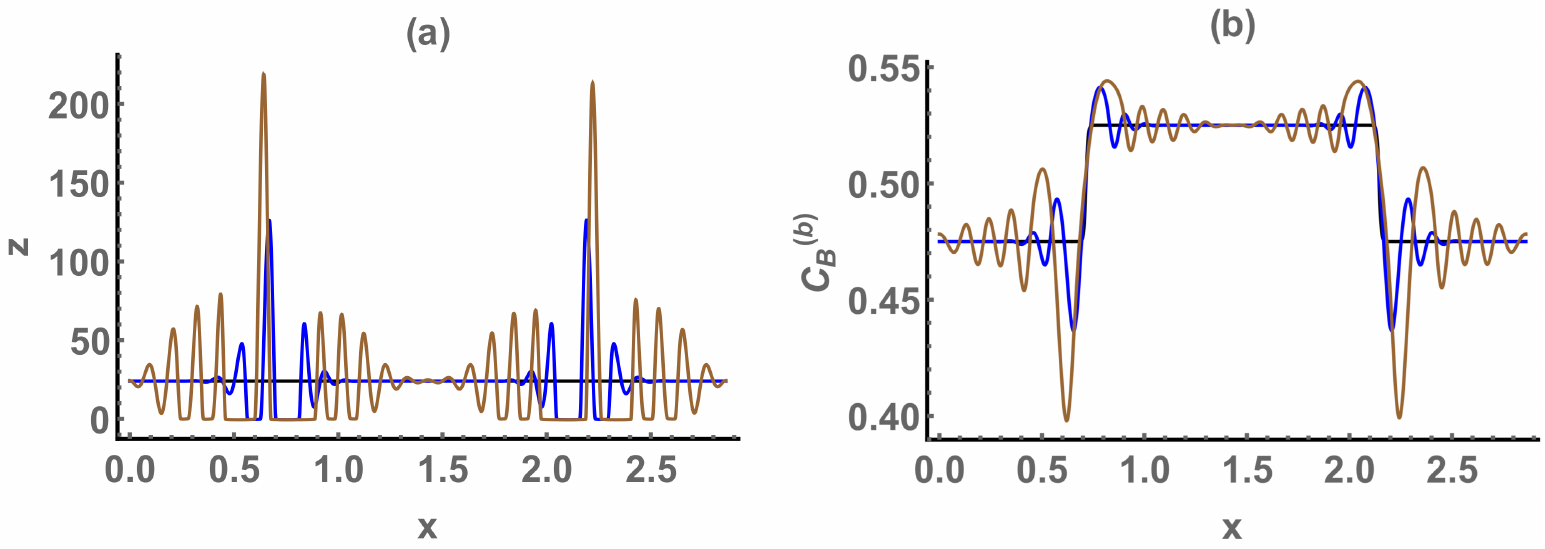}
\vspace{-0.15cm}
\caption{(Color online.) (b) Evolution of the step-like profile of the bulk concentration and (a) the corresponding film morphology (initially the planar surface). 
The final time $t_f=8\times 10^{-5}$ (1 min). 
 }
\label{Fig6}
\end{figure}

In Figures \ref{Fig6}(a,b) the surface morphology starts to evolve first at the locations along the substrate where the initial bulk concentration abruptly changes. At these 
locations form the two largest particles. These particles and the smaller satellite particles are enriched by the A phase.

\begin{figure}[H]
\centering
\includegraphics[width=6.75in]{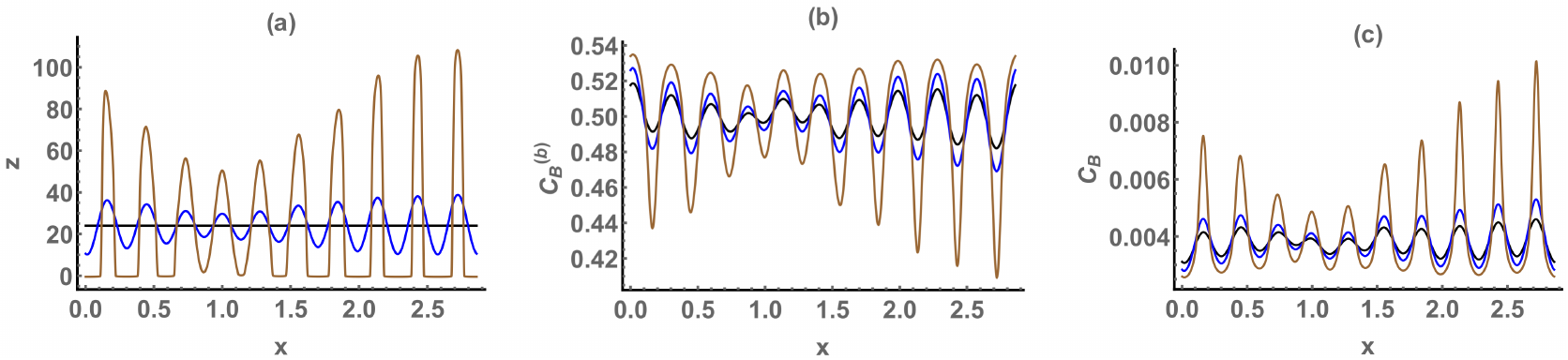}
\vspace{-0.15cm}
\caption{(Color online.) (b) Evolution of the random profile of the bulk concentration, (a) the corresponding film morphology (initially the planar surface) and (c) the
profiles of the surface concentration.
The final time $t_f=5\times 10^{-5}$ (38 s). 
 }
\label{Fig7}
\end{figure}

Lastly, in Fig. \ref{Fig7}(a-c) the superposition of thirteen periodic functions of different periods and amplitudes is taken as the initial condition for the bulk concentration, mimicking a random initial concentration. Here again the largest particles are at the locations where the initial bulk concentration of B atoms is the smallest. In the end of the evolution these particles still have  the smallest $C_B^{(b)}$, but $C_B$ is the largest.

To illustrate the effects of the kinetics on the evolution, Fig. \ref{Fig5} was re-computed first with $M_A=10M_B$ (which is achieved by decreasing the surface diffusivity $D_B$ tenfold from its value in Table \ref{table1}), and then by taking $M_B=10M_A$ in a similar fashion. It must be understood that the kinetic effects here are additive to the thermodynamic ones, as it is not easy to separate them. In both cases the evolution slowed down: the final morphology and the compositions similar to those shown in Fig. \ref{Fig5} were obtained at $t=113$ s and
$t=7400$ s, respectively for two cases. In the latter case the particles had lower content of B atoms (30\% vs. 40\% in Fig. \ref{Fig5}(b)). Unfortunately, 
in the experiment \cite{AKR,AKR1,AR} the kinetic effects were not systematically investigated, thus in regard to these effects our model and the experiment 
can't be compared. Also it is not clear as to whether our results echo the published over-simplified models. For instance, in Ref. \cite{RS} it is found that the
large ratio ($\sim$100) of the surface atomic mobilities for two diffusing species gives the larger linear decay rate of a nanoscale surface ripple. However, the dynamical simulations of a fully nonlinear coupled
PDE system governing the relaxation of a large-amplitude ripple were not performed in this paper, thus the question of the impacts of the kinetic surface segregation 
on the ripple evolution remains partially open. Likewise, in Ref. \cite{KB} the linear growth rate of the surface perturbation is also enhanced when the surface atomic mobilities 
are different, but the long-time nonlinear evolution 
of the morphology is either slightly slowed or sped-up depending on which of the two atomic species has the higher mobility.

We also considered the impacts of $\xi$ by again re-computing Fig. \ref{Fig5} with $\xi=0.1$ and $\xi=10$ (and without the kinetic effects). The morphologies are again similar to Fig. \ref{Fig5}(a).
With $\xi=0.1$ the particles' final bulk concentration $C_B^{(b)}=0.4$, and the final surface concentration $C_B=0.24$ both from the computation and from Fig. \ref{Fig9}(b).
The wetting layer is composed of 90\% B atoms.
With $\xi=10$ the final bulk concentration $C_B^{(b)}=0.45$ (thus the bulk composition does not change significantly, since this value is roughly equal to the initial condition), and the final surface concentration $C_B=0$ both from the computation and from Fig. \ref{Fig9}(b). In this latter case only A atoms are present in the surface layer.
The final times are 900 s and 9 s, respectively for two $\xi$ cases.

Next, we attempted to compute the evolution of the morphology and composition for the short wave instability case shown in Fig. \ref{Fig3}(c). 
Still taking $H(x,0)=H_0=24$, we tried $|\eta|$ in the Gaussian form for $C_B^{(b)}(x,0)$ as large as 0.4, but 
were not able to compute the dewetted film in the reasonable time, likely due to a very small perturbation growth rate, see Fig. \ref{Fig3}(c).
However, the concentration evolves on a faster time scale, and we observed that it quickly returned to the equilibrium value $C_B^{(b)}=0.5$,
even as the film morphology continued to evolve and the surface deviation from the initial height $H_0=24$ was large. Thus unexpectedly, in this case 
the bulk composition stays constant at 50\%(A)$/$50\%(B). This case may be not well-posed, since at $\Gamma_A^{(0)}=0.97$ 
the surface composition $C_B$ is not defined with high accuracy: $0<C_B<0.054$ as is seen from Fig. \ref{Fig9}(e). At smaller $\Gamma_A^{(0)}$, when the instability is still short wave,
we computed the dewetting successfully; in agreement with Fig. \ref{Fig9}(e), $C_B$ values in these cases are near zero.

\emph{If instead of perturbing the initial concentration the film surface is deviated from the planar morphology} (by a Gaussian perturbation or a similar one), 
then the surface pit (bump) will soon evolve into a bump (pit) enriched by the A(B) phase. Thus a particle will form where initially
there was a pit. Such morphological response is the opposite of the one that was noticed in the dewetting experiment on the single-component film \cite{SECS} and in the
corresponding model, Eq. (\ref{eq_h_t_1component}):  the surface pit in the single-component film dewets and the bump evolves into a particle. 

Finally, we remark that fast evolution, i.e. small final times in the computations stem from the somewhat larger surface diffusivities $D_i$ than are expected for a typical metal alloy at 650 $^\circ$C \cite{AKR}.
Decreasing these values by the factor of 100 increases the final times drastically; for instance, the final time in Fig. \ref{Fig5} increases to 19 hrs, which is close to the upper bound
in the experiments that employ the annealing temperature about 650 $^\circ$C. Smaller surface diffusivities (mobilities) do not drastically impact the computed morphologies 
and compositions for the initial conditions that we primarily employed, that is, the perturbation of the uniform composition profile. However, for the alternative initial 
condition, i.e.
the perturbation of the planar film surface at the uniform composition, decreasing the diffusivities results in the more familiar scenario, whereby the small surface pit or bump develop into a larger pit, 
whose deepening ceases when the tip reaches the wetting layer. Despite the differences in the morphological evolution, the composition evolution in this scenario is similar to the 
presented case of larger surface diffusivities, thus the wetting layer below the tip is rich in the B phase, and the particles that form to the left and right of the pit are rich in the A phase.

\section{Summary}

This paper presents the multi-physics and multi-parameter PDE model of a core-shell particle formation by the solid-state dewetting of a binary alloy thin film.
The model is used to study the particles composition, size, and the spatial arrangement (in 1D) as a function of the physical parameters and the initially non-uniform bulk composition of the 
film. The closed-form analytical relation that expresses the thermodynamic surface segregation of one component of a phase-separating binary alloy is derived and analyzed.
Major differences are noticed between the well-studied dewetting of a single-component film and a recently introduced dewetting of a binary alloy film. 
Results of the modeling are in the qualitative agreement
with the experiment \cite{AKR}-\cite{AR}. The model may be extended to include the anisotropy, epitaxial and compositional stresses, and 2D effects.
%

%


\vspace{0.5cm}
\noindent
{\bf ACKNOWLEDGMENTS}

\noindent
This work was supported by QTAG grant from WKU Research Foundation and by NSF KY EPSCoR grant 3200000271-18-069.

\end{document}